\documentclass[sigconf]{acmart}
\AtBeginDocument{%
  }
\usepackage{booktabs}
\usepackage{graphicx}
\usepackage{rotating}
\usepackage{multirow}
\usepackage{pifont}
\usepackage{subcaption}
\usepackage{listings}
\usepackage{xcolor}

\newcounter{listingcounter}

\definecolor{checkgreen}{RGB}{0,170,0}
\definecolor{crossred}{RGB}{200,0,0}
\definecolor{approxorange}{RGB}{220,120,0}

\newcommand{\cmark}{\textcolor{checkgreen}{\ding{52}}}
\newcommand{\xmark}{\textcolor{crossred}{\ding{56}}}

\setcopyright{none}
\renewcommand\footnotetextcopyrightpermission[1]{}
\copyrightyear{2026}
\acmYear{2026}
\setcopyright{cc}
\setcctype{by}
\acmConference[CF '26]{Proceedings of the 23rd ACM International Conference on Computing Frontiers}{May 19--21, 2026}{Catania, Italy}
\acmBooktitle{Proceedings of the 23rd ACM International Conference on Computing Frontiers (CF '26), May 19--21, 2026, Catania, Italy}
\acmDOI{10.1145/3801487.3806067}
\acmISBN{979-8-4007-2568-5/2026/05}

\begin{document}

%\title{Rethinking ISA-Level Matrix Extensions Through HBM Processing-in-Memory}
\title{AME-PIM: Can Memory be Your Next Tensor Accelerator?}
\author{Emanuele Venieri}
\email{emanuele.venieri2@unibo.it}
\affiliation{%
  \institution{University of Bologna}
  \city{Bologna}
  \country{Italy}
}

\author{Simone Manoni}
\email{s.manoni@unibo.it}
\affiliation{%
  \institution{University of Bologna}
  \city{Bologna}
  \country{Italy}
}

\author{Alberto Florian}
\email{alberto.florian@studio.unibo.it}
\affiliation{%
  \institution{University of Bologna}
  \city{Bologna}
  \country{Italy}
}

\author{Jaehyun Park}
\email{jh0416.park@samsung.com}
\affiliation{%
  \institution{Samsung Electronics}
  \city{Hwaseong}
  \country{Republic of Korea}
}

\author{Kyomin Sohn}
\email{kyomin.sohn@samsung.com}
\affiliation{%
  \institution{Samsung Electronics}
  \city{Hwaseong}
  \country{Republic of Korea}
}

\author{Andrea Bartolini}
\email{a.bartolini@unibo.it}
\affiliation{%
  \institution{University of Bologna}
  \city{Bologna}
  \country{Italy}
}

\renewcommand{\shortauthors}{E. Venieri \emph{et al}.}

\begin{abstract}
\textbf{
High Bandwidth Memory with Processing-in-Memory (HBM-PIM) offers an opportunity to reduce data movement by executing computation directly inside memory, but current commercial platforms expose limited instruction sets and require specialized software stacks. In this work, we investigate whether HBM-PIM can serve as a backend for ISA-level matrix acceleration, using the RISC-V Attached Matrix Extension (AME) as a semantic reference. We propose a PEP-based execution model that maps AME element-wise and matrix instructions to HBM-PIM micro-kernels and data instructions in memory operations. Differently from SoA HBM-PIM, we introduce a reduction-free outer-product dataflow that enables accumulation entirely within memory despite the lack of native reduction support. Our approach supports end-to-end execution of element-wise operations, GEMV, and GEMM in PIM mode, minimizing host involvement and off-chip transfers. An experimental evaluation on Samsung Aquabolt-XL shows that AME matrix tile multiplication achieves up to 14.9 GFLOP/s (59.4 FLOP/cycle) on a single HBM pseudo-channel.}
\end{abstract}





\maketitle

\section{Introduction}
The performance of modern processors is increasingly limited not by the availability of arithmetic units, but by the ability to supply them with data \cite{aquabolt}. As workloads in machine learning, scientific computing, and data analytics continue to scale, large working sets and low reuse often push execution beyond the effective capacity of on-chip caches. In these regimes, computation is dominated by off-chip memory traffic, and system efficiency is determined by memory bandwidth and the energy cost of moving data across the memory hierarchy. Taken together, these effects constitute the \emph{von Neumann bottleneck}, where data must continuously flow from memory into the CPU pipeline. This limitation is further amplified in the era of generative AI and foundation models, whose working sets frequently exceed on-chip memory capacity by several orders of magnitude.

%HBM Intro
High Bandwidth Memory (HBM) \cite{JEDEC:2013:HBM} was introduced to alleviate this bottleneck by providing substantially higher bandwidth than conventional DDR-based DRAM \cite{aquabolt, HBM_HC}. By stacking DRAM dies and integrating them within the same package as the host using wide interfaces, HBM exposes much higher I/O parallelism while reducing energy per bit transferred \cite{HBM_HC}. This makes HBM an attractive solution for memory, and it has become a common choice in GPUs and high-performance accelerators. However, even with HBM, emerging workloads can remain memory-limited \cite{jia2018dissecting}, and further scaling memory bandwidth through wider interfaces or faster signalling faces practical constraints in packaging, power delivery, thermals, and signal integrity \cite{hbm_si, HBM_HC, HBM_DRAM}.

% HBM-PIM Intro
Processing-in-Memory (PIM) architectures offer a complementary path forward: rather than only increasing the rate at which data can be moved to compute units, PIM reduces the need for data movement by performing computation close to where the data resides \cite{aquabolt}. In HBM-PIM, lightweight compute logic in the memory stack can execute simple arithmetic kernels directly within the memory device, leveraging internal bank-level parallelism and minimizing host involvement. Commercial HBM-PIM platforms such as Samsung Aquabolt-XL \cite{hbm_pim_hc} demonstrate the practicality of this approach, enabling in-memory execution through JEDEC-compliant host commands without custom HBM memory controllers. Moreover, the use of HBM-PIM proved to be effective in both reducing execution time (2.1$\times$ faster) and power consumption (71\% less) than conventional GPU with HBM memory \cite{PIM-power}.

% HBM-PIM Architectural constraints
Despite these benefits, effectively using today’s HBM-PIM remains difficult due to architectural and software constraints. In particular, the set of natively supported operations is limited, and current commercial HBM-PIM Instruction-Set Architectures (ISAs) do not provide native reduction support \cite{hbm_pim_hc, mpc_wrap}. This is largely because the arithmetic blocks in HBM-PIM are implemented in a memory-oriented logic technology with limited logic density, which constrains the amount and complexity of compute that can be integrated on the HBM logic layer. For matrix and vector workloads, this becomes a critical obstacle: while in-memory Multiply-and-Accumulate (MAC) units can efficiently generate partial products, many common dataflows require reductions across those partial results. Consequently, prior studies typically perform reductions on the host CPU \cite{aquabolt, rnn_hbm} or on an attached accelerator such as an FPGA \cite{mpc_wrap}, increasing data movement and partially undermining the motivation for PIM.

%At the same time, fully realizing the benefits of HBM-PIM requires a non-trivial software stack and programming model. Commercial platforms such as Aquabolt-XL rely on PIM-specific runtime layers, driver support, memory allocation constraints, and careful orchestration of memory requests in order to trigger correct in-memory execution \cite{hbm_pim_hc}. In practice, this forces applications to reason about PIM-reserved memory regions, uncached mappings, physically contiguous allocations, and multi-threaded request generation with explicit synchronization. Such requirements make PIM kernels difficult to express in a portable way and increase the effort needed to integrate HBM-PIM across different host processors and toolchains.

% HBM-PIM Software Integration Challenges
Beyond architectural limits, the interface exposed by commercial HBM-PIM platforms also makes integration challenging. Rather than being invoked through a general architectural mechanism, PIM execution is accessed through device-specific primitives and hand-written microkernels that must be explicitly orchestrated by runtime software \cite{hbm_pim_hc}. As a result, applications must manage PIM execution as a specialized workflow that involves mode transitions, constrained memory placement, and carefully scheduled command sequences, instead of relying on standard compiler and OS abstractions. This accelerator-style programming model complicates portability across host processors and toolchains, limits compiler-driven optimization, and substantially increases the engineering effort required to deploy PIM kernels in real systems.

% Introduce Matrix Extensions and AME
In contrast, modern CPUs increasingly expose matrix acceleration through ISA-level extensions, where tiled GEMM/GEMV-style computation is expressed using well-defined architectural semantics. This includes dedicated register files, with the purpose of containing matrix tiles and direct support to algebraic matrix--matrix operations that uses said tile registers as source operands. Examples include Intel Advanced Matrix Extensions (AMX) \cite{intel_ise_prm_amx} and Arm’s Scalable Matrix Extension (SME) \cite{arm_sme_intro}. These extensions show how matrix workloads can be integrated into the ISA of a processor, enabling compilers and software stacks to target specialized execution engines without relying on device-specific APIs.

% Our Work
Motivated by this perspective, can HBM-PIM be used as a backend for CPU ISA extensions for matrix operations, with the goal of enabling more general and portable exploitation of PIM through well-defined architectural semantics? As a semantic reference point, we use the T-Head proposal for the RISC-V Attached Matrix Extension (AME) \cite{t_head_ame_v0_6_0}, which provides an ISA-level abstraction for tiled matrix execution that could be naturally exposed to compilers and software stacks. Rather than treating HBM-PIM as a specialized memory device accessed through custom APIs, we study how its compute capabilities can be mapped onto matrix-extension-style operations in a way that preserves clean semantics while respecting real platform constraints.

We propose an outer-product--based mapping that remaps matrix execution to exploit in-memory MAC units for accumulation directly within HBM, avoiding explicit reduction operations that are unsupported in current HBM-PIM platforms. Previous works \cite{mpc_wrap,rnn_hbm} relied on external reduction engines to overcome this limitation, while our approach enables element-wise operations as well as GEMV and GEMM to execute entirely inside HBM-PIM, reducing host interaction and minimizing off-chip transfers. Our results show that carefully designed dataflows can significantly improve the practicality of HBM-PIM for ISA-level acceleration, and provide insights for future PIM architectures and programming models.

Overall, this paper makes the following contributions:
 \vspace{-0.8mm}
\begin{itemize}
    \item We study how commercial HBM-PIM devices can serve as a practical backend for matrix ISA extensions subsets, using AME semantics as a reference abstraction.
    \item We introduce an outer-product--based execution strategy that performs accumulation within HBM via in-memory MAC units, avoiding host-side reduction phases.
    \item We experimentally evaluate the proposed mapping on Samsung Aquabolt-XL. Our results show that AME GEMM, GEMV, and element-wise instructions can execute fully in PIM mode, reaching 14.9 GFLOP/s (59.4 FLOP/cycle) on a single pseudo-channel for the mfmacc instruction on 128 $\times$ 4096 tiles. %On a single pseudo-channel, matrix tile multiplication reaches up to 14.9 GFLOP/s, achieving 59.4 FLOP/cycle for the mfmacc instruction on 128 $\times$ 4096 tiles, with clear scaling trends as tile sizes increase.
    \item We identify key ISA and microarchitectural limitations --- missing reduction support and restricted operand routing ---and show how they constrain AME execution efficiency, informing the design of future HBM-PIM architectures. %of current HBM-PIM platforms—such as missing reduction support and restricted operand routing—and show how they affect the efficiency of AME-style matrix execution, informing the design of future HBM-PIM architectures with richer semantics.
\end{itemize}
\section{Background}
This section summarizes the organization and programming interface of Samsung’s Aquabolt-XL HBM-PIM platform, which serves as the commercial baseline for our work. We focus on the architectural features most relevant to matrix and vector execution, including the placement and structure of the in-memory compute units, the operating modes used to expose bank-level parallelism, and the ISA through which PIM microkernels are expressed and executed.

\subsection{HBM-PIM Hardware Architecture}

\begin{figure}[t]
    \centering
    \includegraphics[width=\linewidth]{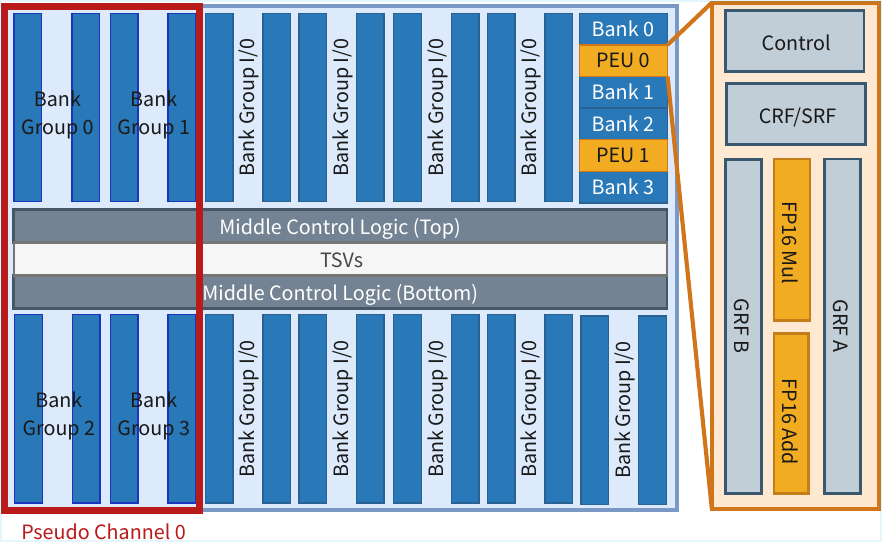}
    \caption{Samsung Aquabolt-XL PIM DRAM architecture}
    \label{fig:hbm_pim_arch}
\end{figure}

Aquabolt-XL augments an HBM2-based stacked memory system with in-memory compute capability while preserving conventional DRAM operation. Rather than modifying DRAM subarrays, PIM functionality is implemented by inserting a lightweight \emph{PIM unit} at the I/O boundary between the DRAM cells and the peripheral circuitry. Each PIM unit is shared by a pair of DRAM banks, referred to as the even and odd bank, and can access one of the two banks at a time. This design choice reduces area and power overhead while maintaining direct access to bank-local data.

Each PIM unit consists of three components: a 16-wide SIMD floating-point unit, a set of local register files, and a controller. The SIMD datapath comprises sixteen FP16 lanes, each equipped with a multiplier and an adder, organized as a five-stage pipeline. The register files include a Command Register File (CRF) with 32 entries of 32 bits each, a General Register File (GRF) with sixteen 256-bit registers split between the even (GRF\_A) and odd banks (GRF\_B), and a Scalar Register File (SRF) composed of two banks of eight registers for scalar multiplication (SRF\_M) and addition (SRF\_A). The controller fetches and decodes PIM instructions from the CRF and orchestrates operand selection, SIMD execution, and write-back to the GRFs.
The HBM stack consists of four DRAM dies and four PIM-enabled DRAM dies. Each PIM-enabled DRAM die is divided into four pseudo-channels, each connected to sixteen DRAM banks. With one PIM unit shared between each pair of banks, each pseudo-channel integrates eight PIM units. Figure \ref{fig:hbm_pim_arch} shows the internal architecture of an Aquabolt-XL HBM-PIM die.
All PIM units operate at the same frequency as the DRAM core, between 250\,MHz and 300\,MHz, and collectively provide substantial on-chip compute bandwidth for vector and matrix operations.

From the host perspective, Aquabolt-XL is accessed using standard HBM transactions. In PIM mode, DRAM column read and write commands are repurposed to trigger PIM unit execution with deterministic latency. This command-driven execution model preserves compatibility with JEDEC-compliant memory controllers while enabling tightly synchronized, high-throughput execution across multiple banks inside the HBM stack.

\subsection{Operating Modes}
For each pseudo-channel Aquabolt-XL HBM-PIM supports three operating modes:(i) \emph{Single-Bank (SB) Mode} corresponds to standard HBM behavior, where an Activate (ACT) command selects a specific bank and subsequent column commands to that bank access its open row. This mode is used for conventional DRAM memory accesses.
(ii) \emph{All-Bank (AB) Mode} overrides bank address fields so that a single column command targets the same row and column across all banks in a pseudo-channel simultaneously. This exposes bank-level parallelism in lock-step fashion and is also used to broadcast microkernel instructions into the CRF of every PIM unit during the programming phase. 
%To increase internal parallelism, Aquabolt-XL supports an all-bank (AB) access mode in which the bank address fields are effectively overridden. In this mode, a single-column command targets the same row and column across all banks within a pseudo-channel, enabling multiple banks to be accessed simultaneously. As a result, AB mode increases the amount of data accessed per command and exposes bank-level parallelism in a structured, lock-step fashion. In addition to multiple-banks data access, AB mode is employed to broadcast microkernel instructions during PIM programming , allowing the same instruction sequence to be loaded into the CRF of every PIM unit within the selected pseudo-channel.
(iii) \emph{All-Bank PIM (AB-PIM) Mode}\label{par:ab-pim}  builds on all-bank (AB) mode by coupling each DRAM column command with in-memory execution. %When AB-PIM mode is enabled, a column command no longer serves solely as a data access, but instead triggers the PIM execution units attached to the banks to fetch and execute an instruction from the CRF. 
Each such command advances an internal PIM program counter, causing all PIM units within a pseudo-channel to step through a preloaded microkernel in lock-step, while also acting as a pointer to the bank data used as a source operand. % The issued column command also acts as pointer to bank data used as source operand in PIM instructions. 
(iv) \emph{Mode transitions} are performed via predefined DRAM command patterns to a reserved configuration memory region, preserving compatibility with unmodified JEDEC-compliant controllers.

\subsection{PIM Instruction Set Architecture}
Aquabolt-XL exposes a compact instruction set designed to support memory-bound vector and matrix kernels directly within the HBM stack. PIM instructions are executed by the PIM units placed at the bank I/O boundary and are stored in a per-pseudo-channel CRF. Each instruction is encoded in a fixed 32-bit format and is fetched and executed in response to DRAM column commands when operating in AB-PIM mode.
The PIM instruction set is intentionally minimal and consists of three main classes: arithmetic operations, data movement operations, and control-flow instructions.

\subsubsection{Arithmetic Instructions}
Inside each PIM unit, arithmetic instructions operate on the 16-lane FP16 datapath and are optimized for vector-vector and vector-matrix computation. The supported operations include \texttt{ADD}, \texttt{MUL}, \texttt{MAD}, and \texttt{MAC}, which together enable fused multiply-add style computation commonly used in linear algebra and machine learning workloads. In particular, the \texttt{MAC} instruction accumulates the product of two operands into a destination register, making it the primary building block for GEMV- and GEMM-like kernels. There is no availability of cross-lane reduction-like instructions.

Operands for arithmetic instructions can be sourced from multiple locations, including the GRFs, SRFs, or directly from the active DRAM row buffer associated with a bank. Results are written back to the GRF or to bank-local storage, allowing accumulation to remain entirely within the memory device. To further increase efficiency, Aquabolt-XL supports an address-aligned mode (AAM) in which operand selection is implicitly derived from the command's DRAM row and column addresses, enabling multiple arithmetic operations to be issued with a small number of instructions.

\subsubsection{Data Movement Instructions}
Data movement instructions are used to transfer data between the DRAM row buffer and the PIM-local registers. The instruction set provides \texttt{MOV} and \texttt{FILL} operations, which load data from banks into registers or write register contents back to memory. These instructions support multiple operand routing configurations, allowing flexible movement between bank storage, GRFs and SRFs. Broadcasts of data from a single bank to all GRFs of a pseudo-channel is also possible.

Certain data movement operations can optionally apply simple activation functions such as ReLU as part of the transfer, further reducing the need for additional compute stages. However, all data movement remains explicitly programmed and tightly coupled to the underlying DRAM access pattern.  In this work we do not leverage this capability as the AME specification does not have dedicated instructions for activation functions on matrix tiles. Future works can combine the AME-PIM with Neural Networks compilers/Design Specific Architecture to leverage it.

\subsubsection{Control-Flow Instructions}
To support repetitive kernel execution, the PIM ISA includes a small set of control-flow instructions, namely \texttt{NOP}, \texttt{JUMP}, and \texttt{EXIT}. These instructions allow microkernels stored in the CRF to implement loops and termination without host intervention. In particular, \texttt{JUMP} instructions are optimized for zero-cycle execution by relying on predecoded iteration counts, enabling efficient looping despite the absence of a full-fledged control unit. This approach has the drawbacks not allowing \texttt{JUMP} nesting and limiting the loop counter to a maximum of 255 iterations. 

\subsubsection{Implications for Matrix Execution}
Overall, the PIM instruction set is sufficient to express the core arithmetic patterns of matrix and vector workloads, especially those dominated by multiply-accumulate operations. At the same time, it lacks support for more complex operations such as comparisons, reductions, and data-type widening, which constrains the class of computations that can be executed entirely within memory. These characteristics directly influence how higher-level matrix abstractions can be mapped onto HBM-PIM and motivate the execution strategies we explored in this work.
\vspace*{-5pt}
\subsection{AME extension by T-Head}
To provide an ISA-level abstraction for tiled matrix computation, we rely on the AME proposal published by T-Head \cite{t_head_ame_v0_6_0}. It defines a complete programmer-visible model for an attached matrix extension, including architectural state, configuration registers, tile-level and element-wise instruction semantics. We leverage this proposal as our reference specification since, despite not being ratified yet, it has already been used as a foundation in prior works \cite{quadrilatero, xuantie_c907_ame}.
\vspace*{-5pt}
\subsubsection{Architectural state and configuration}
AME introduces dedicated architectural resources for matrix execution in addition to the RISC-V scalar register file. The extension defines two groups of matrix registers: a set of \emph{tile registers} (\texttt{tr0--tr3}) used to hold input tiles, and a set of \emph{accumulation registers} (\texttt{acc0--acc3}) used to store partial sums and output tiles.

To support flexible tiling, AME includes configuration state that determines the logical tile dimensions used by subsequent instructions. In particular, the proposal defines configuration Control and Status Registers (CSRs) controlling the active matrix shape through the parameters \(M\), \(K\), and \(N\) (e.g., \texttt{mtilem}, \texttt{mtilek}, \texttt{mtilen}), allowing software to adapt tile sizes at runtime for boundary conditions or problem-specific blocking.
\vspace*{-3pt}
\subsubsection{Instruction set overview}
The T-Head AME proposal organizes its instruction set into five groups. % \emph{configuration}, \emph{matrix multiplication}, \emph{load/store}, \emph{Misc}, and \emph{element-wise} instructions.\\
\emph{Configuration instructions} set the active tile dimensions \(M\), \(K\), and \(N\) via \texttt{msettile\{m,k,n\}}-style mechanisms and provide lifecycle management via \texttt{mrelease}.
\emph{Matrix multiplication instructions} define the core tiled compute primitive: a multiply-and-accumulate that consumes two source tiles from \texttt{tr} registers and accumulates results into an \texttt{acc} register. It supports multiple operand/accumulator datatype combinations, including floating-point and integer dot-product variants, as well as hybrid-precision/widening forms depending on the supported implementation subset.
\emph{Load/store instructions} move matrix tiles between memory and matrix registers. These instructions explicitly encode the element size and tile layout, and include variants to support access patterns such as whole-register transfers and transposed forms, which are commonly required to implement blocked GEMM/GEMV kernels efficiently.
\emph{Misc instructions} cover auxiliary functionality %needed to compose tiled kernels, including
such as data movement between matrix registers, scalar--matrix moves, broadcast operations, packing, and slide operations over rows/columns.
\emph{element-wise instructions} operate per element of a tile register and include arithmetic (add, subtract, multiply)% operations such as addition, subtraction, multiplication, as well as 
comparison-based (min, max), shifts, and data type conversions.

Overall, AME provides a tile-based ISA abstraction in which matrix computation is expressed in terms of explicit architectural registers, configurable tile dimensions, and structured tile-level memory and arithmetic operations.

\section{Methodology}
{%  <-- open a group so the rename is local
\renewcommand{\figurename}{Listing}
\refstepcounter{listingcounter}
\renewcommand{\thefigure}{\arabic{listingcounter}}
\begin{figure*}[t]
    \centering
    \subfloat[ADD(MUL)-PEP listing.\label{lst:add-mul-pep-lst}]{%
        \includegraphics[width=0.32\textwidth]{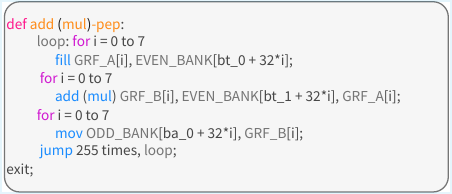}}
    \hfill
    \subfloat[SUB-PEP listing.\label{lst:sub-pep-lst}]{%
        \includegraphics[width=0.32\textwidth]{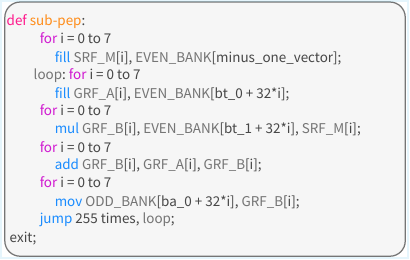}}
    \hfill
    \subfloat[MAC-PEP listing.\label{lst:mac-pep-lst}]{%
        \includegraphics[width=0.32\textwidth]{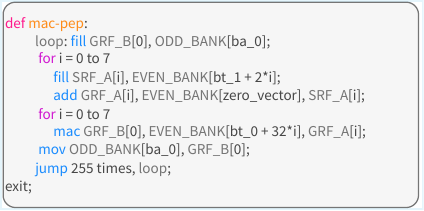}}
    \caption{Pseudocode listings of the four proposed PEPs. ADD-PEP and MUL-PEP are presented
    as a single listing for brevity, since their implementations differ only in the core PIM
    arithmetic instruction.}
    \label{lst:pep-listings}
\end{figure*}
}

This subsection describes how in our work the AME programming model is realized on the Aquabolt-XL HBM-PIM platform. It is structured as follows. Section \ref{bridge} analyzes which AME computational primitives can be directly mapped onto the native PIM instruction set and identifies the architectural constraints that limit full ISA coverage. Section \ref{ame_in_hbm} presents our implementation strategy in which AME tile and accumulation registers are realized as memory-resident abstractions within HBM-PIM and their execution is driven by PIM execution primitives. Section \ref{progmodel} discusses the realization of AME control and status registers and the exposed programming interface, completing the description of a PIM-based AME implementation.

%\subsection{Bridging T-HEAD to PIM ISA} 
\subsection{AME-PIM ISA comparison} 
\label{bridge}
We first evaluate whether AME arithmetic and matrix instructions can be executed natively within HBM-PIM, without relying on external computational hardware.

Table \ref{tab:ame-pim-instr} summarizes the mapping between individual AME instructions and the corresponding PIM instructions used for their implementation. In most cases, an association is feasible. However, exceptions arise for operations involving widening, as the PIM datapath supports only the FP16 format, and for minimum and maximum operations, due to the absence of conditional or comparison instructions in the PIM execution model.
\begin{table}[h!]
\begin{tabular}{l|l}
AME instruction        & PIM instructions \\ \hline
mfadd.h.mm             & ADD            \\
mfadd.h.mv.i           & ADD            \\
mfsub.h.mm             & MUL, ADD       \\
mfsub.h.mv.i           & MUL, ADD       \\
mfmul.h.mm             & MUL            \\
mfmul.h.mv.i           & MUL            \\
mfmax.h.mm             & Not supported  \\
mfmax.h.mv.i           & Not supported  \\
mfmin.h.mm             & Not supported  \\
mfmin.h.mv.i           & Not supported  \\
mfmacc.h               & MAC            \\
mfmacc (with widening) & Not supported 
\end{tabular}
\vspace{3mm}
\caption{Summary of the association of AME arithmetic element-wise and matrix multiplication instructions with PIM instructions.}
\label{tab:ame-pim-instr}
 \vspace{-7mm}
\end{table}

\subsection{AME-PIM Implementation} \label{ame_in_hbm}
In this paper we propose small computational kernels, called PIM Execution Primitive (PEP), to implement AME instructions using the PIM ISA. The PEPs encapsulate sequences of native PIM instructions that are executed entirely within the HBM-PIM. The defined PEPs are used iteratively and in combination with tile register memory mapping we adopted to implement AME instructions. 

The PEPs employ a dataflow similar to the AME execution model, transferring data from Even Banks to Odd Banks through internal PIM operations involving the GRFs. This organization aligns with AME requirements for tile and accumulation registers. In this work we propose not to physically instantiate these registers but to abstract them as memory-resident data structures accessible by the PIM units as described in the previous paragraph. Even Banks store data corresponding to AME tile registers, while accumulation registers are mapped to Odd Banks. 

\subsubsection{Tile Registers Memory Layout} \label{tile_in_hbm}
The organization of matrix tiles in memory is dictated by the execution model of HBM-PIM. Each PIM unit operates on 16 elements in parallel, and PEPs are executed concurrently across the eight PIM units within a pseudo-channel. Moreover, the absence of cross-lane reduction requires each SIMD lane to process elements from the same row of a tile.
To satisfy these constraints, we map tile rows across the Even Banks of a pseudo-channel and store data in column-major order within each bank. This layout ensures that contiguous memory accesses correspond to elements belonging to a single row, aligning with the SIMD execution requirements.
As a consequence, the number of rows processed in parallel is fixed by the number of Even Banks, resulting in 128 rows per tile. This fixed row size is a fundamental constraint that drives our implementation of AME matrix tiles.

\begin{figure}[t!]
    \centering
    \includegraphics[width=\linewidth]{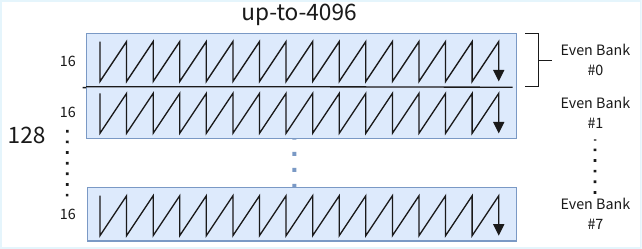}
    \caption{Tile register memory mapping in the Even Banks of a pseudo-channel.}
    \label{fig:tile-map}
\end{figure}

The maximum number of columns of a Tile register is then derived from the number of Tile registers required by AME specification, in this case equal to four, and the available storage of the Even Banks.
From HBM-PIM specification each pseudo channel has available 1 Gb of storage. Considering that this is halved in Even and Odd banks, then the total amount of space available allows the presence of 4 Tile registers of up to 32768 columns. However, this exceeds the maximum dimension supported by the AME specification, which for FP16 data is limited to 4,096 elements per row. Accordingly, in our implementation we restrict tile dimensions to $128\times4096$. In Figure \ref{fig:tile-map} it is represented this matrix tile mapping inside a pseudo-channel. 

\begin{figure}[t!]
    \centering
    \includegraphics[width=0.9\linewidth]{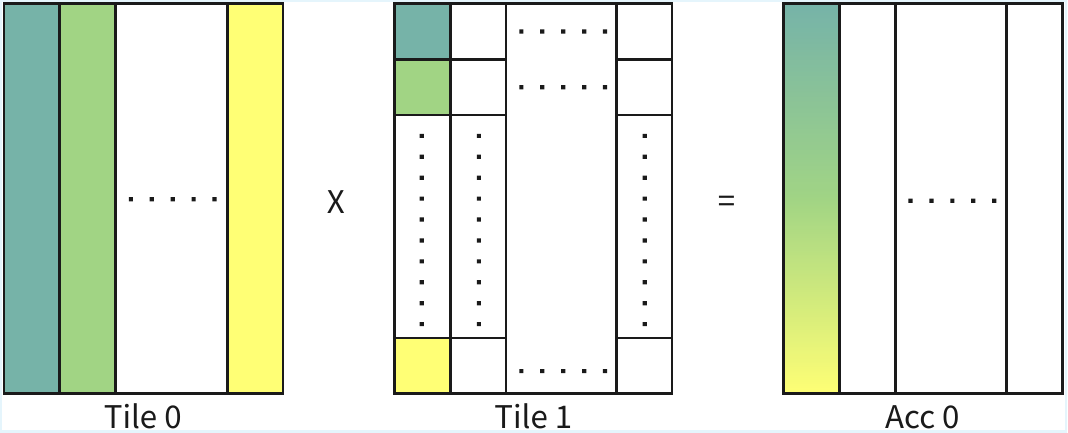}
    \caption{Reduction-free matrix multiplication via MAC-PEP: outer-product decomposition through in-memory vector–scalar MACs.}
    \label{fig:vec-scalar}
    \vspace{-4mm}
\end{figure}

\subsubsection{AME instructions mapping onto PIM}
\begin{figure*}[t]
    \centering
    \includegraphics[width=\textwidth]{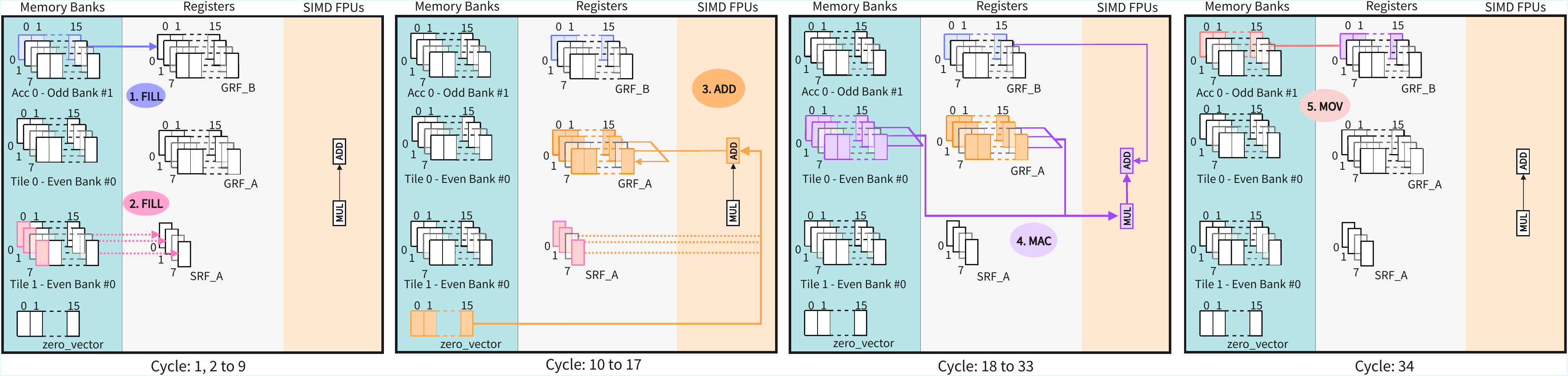}
    \caption{Cycle-by-cyle execution of the \emph{mfmacc} AME instruction through the MAC-PEP, illustrating the dataflow across memory banks, PIM registers (GRF, SRF), and SIMD FPUs over the five micro-operations.}
    \label{fig:mac-pep-exec}
\end{figure*}
On HBM-PIM, each PEP is a binary code of PIM instructions loaded in the CRF of each PIM unit during the setup phase. In the listings showed in Listing~\ref{lst:pep-listings}, operations are expressed in pseudo-code, with each operand representing a block of 16 FP16 values processed in a SIMD fashion. Execution proceeds in AB-PIM Mode as described in Section \ref{par:ab-pim}, addressing operands from a base address plus an offset (in Listings indicated as \texttt{bt\_\#} or \texttt{ba\_\#}, for the base of tiles and accumulators respectively). In general, FP16 instructions relies on AAM, allowing eight iterations of the same instruction on consecutive eight blocks of FP16 values. In the listings this is reported as \texttt{for} loop with a counter that increments from 0 to 7. This behaviour is enabled or disabled in the instructions themselves via a dedicated bit-field.

\subsubsection{mfadd/mfmul instructions}
The \texttt{ADD-PEP} and \texttt{MUL-PEP} (Listing~\ref{lst:add-mul-pep-lst}) implement the AME addition and multiplication instructions on two $128 \times 2048$ tiles, iterating 256 times on $128 \times 8$ windows and writing results to the Odd Banks. Smaller column dimension are supported by reducing the \texttt{JUMP} instruction \emph{loop counter}, at the cost of increased relative setup overhead; smaller row dimensions result in unused SIMD lanes, as the parallel execution width is fixed by the PIM architecture.

%The AME addition and multiplication instructions are implemented using the \texttt{ADD-PEP} and \texttt{MUL-PEP} (Listing \ref{lst:add-pep}), which natively operates on two $128 \times 2048$ tiles, iterating 256 times on windows of $128 \times 8$ data, storing the results and moving on the next window. Smaller tiles along the column dimension can be supported by reducing the \texttt{JUMP} instruction \emph{loop counter}, while smaller row dimensions are handled by simply ignoring computed results in the exceeding dimension.
%In the former case, the resulting \texttt{FLOP/cycle} decreases because the fixed setup overhead becomes more significant relative to the reduced execution time. In the latter case, computational resources remain underutilized due to the rigid parallel execution pattern imposed by the PIM architecture.

\subsubsection{mfsub instruction}
Subtraction is not natively supported by the PIM ISA, and it is emulated by multiplying the subtrahend by \texttt{-1}
using the \texttt{MUL} instruction with the \texttt{SRF\_Ms} scalar broadcast register, then adding the result to the minuend. This requires reserving a small portion of the Even Bank to store a vector of \texttt{-1} values for the \texttt{SRF\_Ms} initialization. The resulting \texttt{SUB-PEP} (Listing~\ref{lst:sub-pep-lst}), operates identically to the other PEPs with a moderate throughput reduction due to the additional instructions required.
\begin{table}[t]
\begin{tabular}{l|l}
AME implementation constant & Value \\ \hline
ELEN  & 16 bits\\
TLEN & $2^{23}$ bits\\
TRLEN & $2^{16}$ bits\\
ROWNUM & 128\\
ARLEN & $2^{16}$ bits\\
ALEN & $2^{23}$ bits
\end{tabular}
\caption{Summary of the implementation-defined constant parameters values}
\label{tab:ame-consts}
\vspace{-8mm}
\end{table}
\subsubsection{mfmacc instruction}
The proposed \texttt{MAC-PEP} (Listing~\ref{lst:mac-pep-lst}), in conjunction with the memory mapping of matrix tiles described in Paragraph~\ref{tile_in_hbm}, enables the complete execution of matrix multiplication within HBM-PIM without relying on external resources such as reduction engines, as required by previously proposed solutions~\cite{mpc_wrap}.
The computation follows an outer-product formulation implemented through a sequence of matrix--vector operations. At each step, a column of the first source tile is multiplied by a scalar element from the second source tile, and the resulting vector is accumulated into the corresponding column of the destination accumulator tile. 
We leverage here a double broadcasting of the second source matrix operand, from a memory location to all PIM units \texttt{SRF\_As}, then from each of them to GRF\_As using an \texttt{ADD} with a vector of zeros. This requires two instructions in the current PIM ISA. This is due to the impossibility of using \texttt{SRF\_Ms} as input operand for PIM \texttt{MAC} instruction in Address Aligned Mode (AAM).
As a result, matrix multiplication can be performed entirely within the native data movement and execution model of the HBM-PIM. 

The sequence of operations is depicted in Figure~\ref{fig:mac-pep-exec}, which shows the different execution phases as a timeline of events across the HBM-PIM micro-architecture relative to \texttt{PIM Unit \#0}, with each phase replicated separately for clarity.
%The MAC-PEP defined for the \emph{mfmacc} instruction is flexible in terms of source operands as it works regardless of the shape of the tile. However, the computational efficiency depends on it as it will be detailed in section \ref{eval}.
%A single execution of a \texttt{MAC-PEP} can operate on a $128\times2048\times1$ matrix--vector multiplication or a $128\times8\times256$ matrix--matrix multiplication. Size is limited by the \texttt{JUMP} instruction maximum repetition, equal to 256.  

The \texttt{MAC-PEP} operates independently of the tile shape, supporting both $128\times2048\times1$ matrix--vector and $128\times8\times256$ matrix--matrix multiplications within a single execution, with the tile size bounded by the \texttt{JUMP} instruction's maximum iteration count of 256. The impact of tile shape on computational efficiency is discussed in Section~\ref{eval}.
%To fully implement the semantics of the AME \emph{mfmacc} instruction, an initial step populates \texttt{GRF\_B} from the Odd Banks, thereby initializing the accumulator register used as an input operand for the computation, as illustrated in line~3 of Listing~\ref{lst:mac-pep}.
\vspace{-1mm}
\subsubsection{Data movement instructions}
As previously described, in this work we propose to map the AME ISA directly onto the HBM-PIM computational primitives with AME tiles and accumulation registers abstracted as memory regions in the HBM-PIM DRAM dies. In addition to computational instructions, AME defines dedicated \emph{load} and \emph{store} operations, as well as instructions for matrix register management. We implemented these in our proposed approach by means of a hardware table that tracks the memory addresses associated with each AME tile and accumulation register. Specifically, for each \texttt{tr0--tr3} and \texttt{acc0--acc3}, the table records the corresponding memory base addresses. This indirection enables the translation of operand fields in issued AME instructions into the appropriate memory transactions required by the PIM units to operate on the correct data.

Instructions that write, reorganize, or transform data --- such as transposed load, store, move, broadcast, pack, and slide --- are implemented as memory operations that produce new memory locations. Upon completion, the hardware table is updated to associate the affected AME registers with the newly generated locations, while the requested data transformation is carried out within memory.

It has to be noted that data movement instructions in AME are very important, because they enable packing and unpacking routines, which pre-process data to leverage cache locality. In our proposal these instructions can be resolved without a real data movement, but by just updating the pointers in the hardware table. 

\subsection{Programming Model} \label{progmodel}
The AME-PIM implementation fully supports the CSRs, which expose configuration, synchronization, and execution state to software in a way that is consistent with the original proposal. Table \ref{tab:ame-consts} summarizes the maximum implementation-defined constant AME parameters based on the proposed implementation. Tile operands are memory-resident data structures within HBM-PIM, with a maximum supported dimension of $128 \times 4096$. The \emph{msettile{m|k|n}[i]} configuration instructions update both the CSR state and the PEP loop counters, ensuring that tile dimension changes are correctly reflected in all subsequent PIM kernel executions.
%\vspace{-7mm}
%Tile operands are data structures within the HBM-PIM, with a maximum supported tile dimension of $128 \times 4096$. Their actual size can be controlled with the \emph{msettile{m|k|n}[i]} instruction and besides, changing the \texttt{m,k,n} dimensions of the tiles inside the dedicated CSRs, this actually influences PEPs execution allowing the perfect implementation of the corresponded AME instruction. 
\section{Results} \label{eval}
This section presents the experimental results, providing an initial assessment of the performance of an AME implementation based on HBM-PIM.
\begin{figure}[t!]
    \centering
    \includegraphics[width=\linewidth]{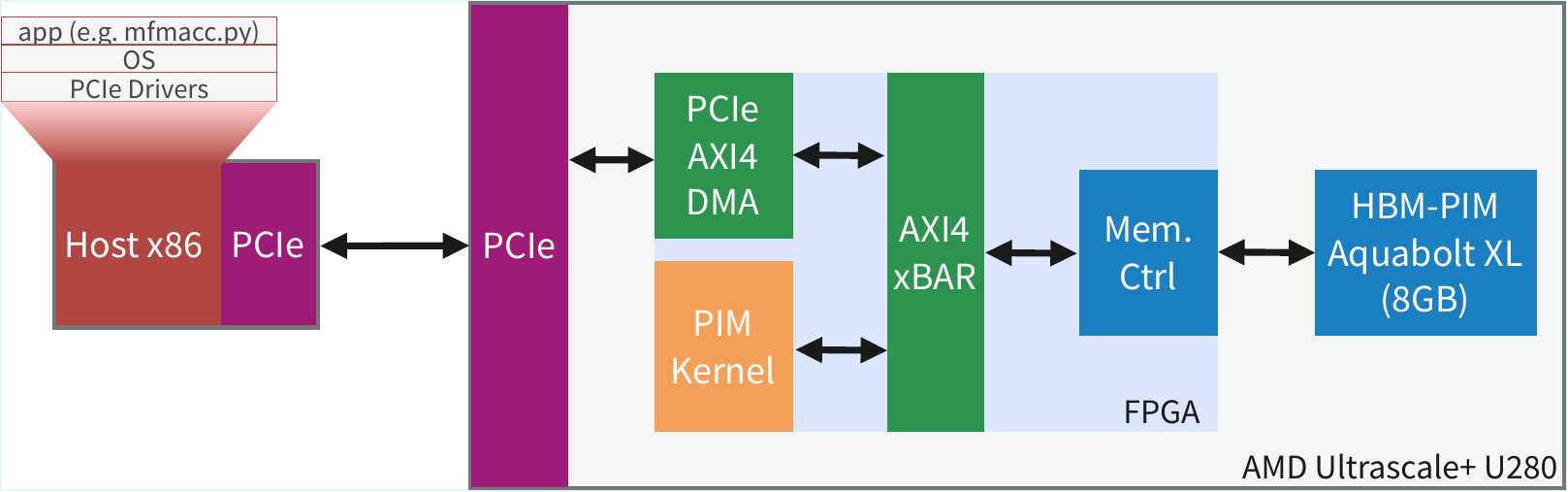}
    \caption{Evaluation platform.}
    \label{fig:eval-setup}
\end{figure}

\subsection{Experimental Setup}
The evaluation was conducted on a Xilinx Alveo U280 FPGA equipped with Samsung Aquabolt-XL HBM-PIM, connected to an x86 host processor via PCIe. The FPGA fabric hosts a simple logic design comprising a memory controller, a PCIe DMA IP, an AXI4 crossbar, and a dedicated encrypted PIM management block that issues PIM commands, referred to as \texttt{PIM\_kernel}, as illustrated in Figure~\ref{fig:eval-setup}. AME instructions are emulated through host-side applications that issue commands to the HBM-PIM via the PCIe driver, triggering execution on the FPGA. This setup is adopted purely for evaluation purposes, as
we do not envision AME-PIM as a PCIe accelerator. The \texttt{PIM\_kernel} block coordinates execution within a single pseudo-channel but is designed around a fixed access pattern for matrix--vector multiplications, which limits the set of programs that can be directly evaluated~\cite{mpc_wrap}.

In the following evaluation we characterized both the execution cycles as well as the setup cycles used for programming the PEP kernel and configuring the HBM-PIM. % The evaluation focuses on two main components: the number of cycles required to configure the PIM for execution and the number of cycles consumed by the execution itself. 
Cycle counts were collected using FPGA-mapped hardware counters triggered by the specific memory transactions that enable PIM functionality. As performance counters are not directly available within the HBM-PIM, the reported measurements reflect performance observed from the bus side. As reported by Lee et al. \cite{aquabolt} the FPGA bus frequency of 250 MHz matches the PIM units one.

\subsection{PEP and AME Instructions Characterization}
\begin{figure}[t]
    \centering
    \includegraphics[width=\linewidth]{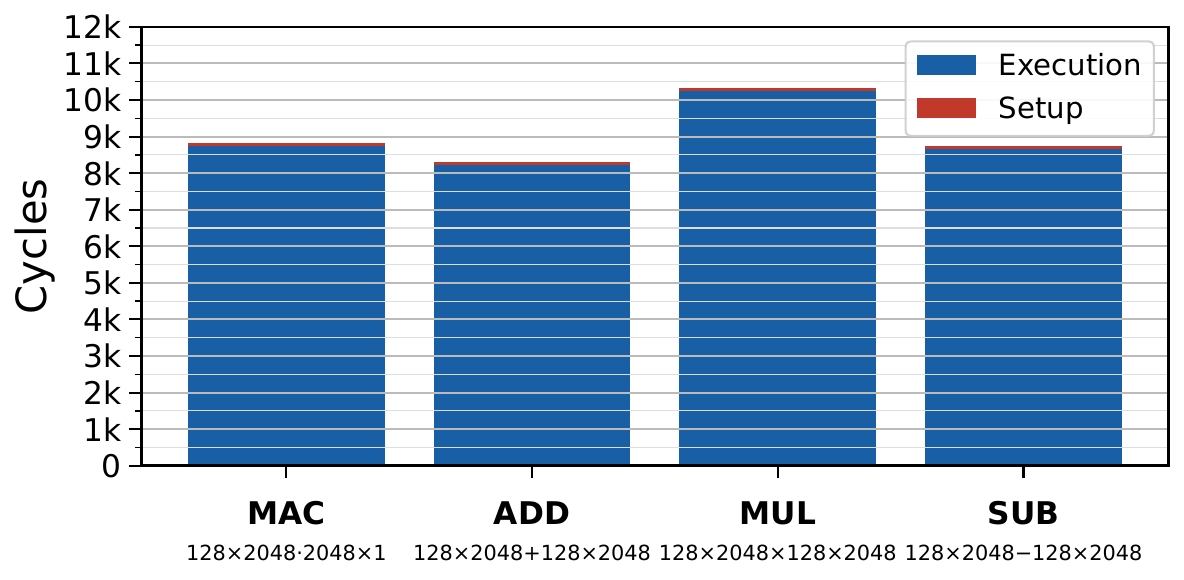}
    \caption{Cycle counts measured for PIM Execution Primitives. Each primitive is annotated with operand dimensions.}
    \label{fig:pep_cycles}
\end{figure}
In this subsection, we evaluate the cycles associated with the execution of individual PEPs and performance of AME instructions implemented via those. These measurements are reported in Figure~\ref{fig:pep_cycles} and exhibit limited variability across different operations. This behaviour is primarily attributable to the constraints of the available HBM-PIM experimental platform, in which PIM trigger commands are generated by PIM\_kernel implemented on the FPGA, resulting in a largely uniform execution latency across PEP types.
\begin{figure}[t]
\vspace{-2.2mm}
    \centering
    \subfloat[Cycles of AME instructions\label{fig:ame-instr-cycles}]{%
        \includegraphics[width=0.49\linewidth]{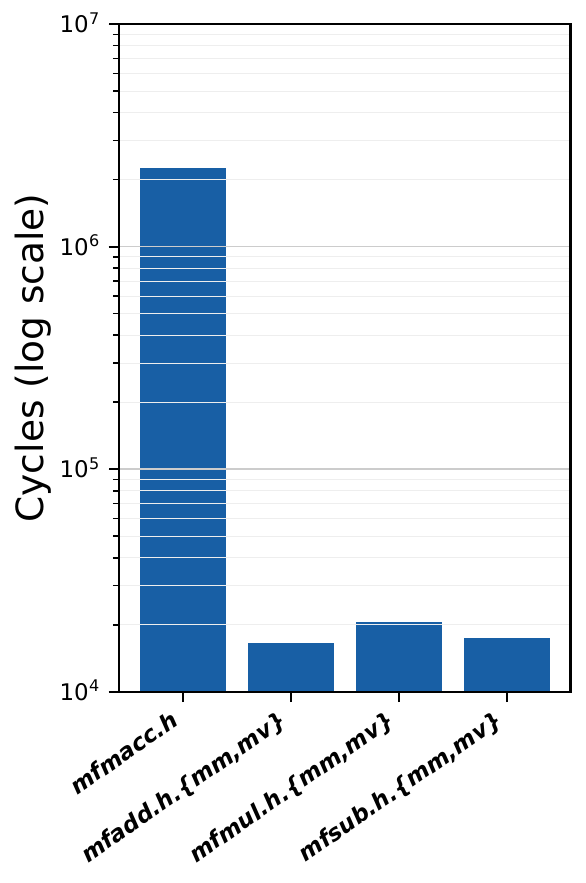}}
    \hfill
    \subfloat[FLOP/Cycle and GFLOPS AME instructions\label{fig:ame-instr-gflops}]{%
        \includegraphics[width=0.49\linewidth]{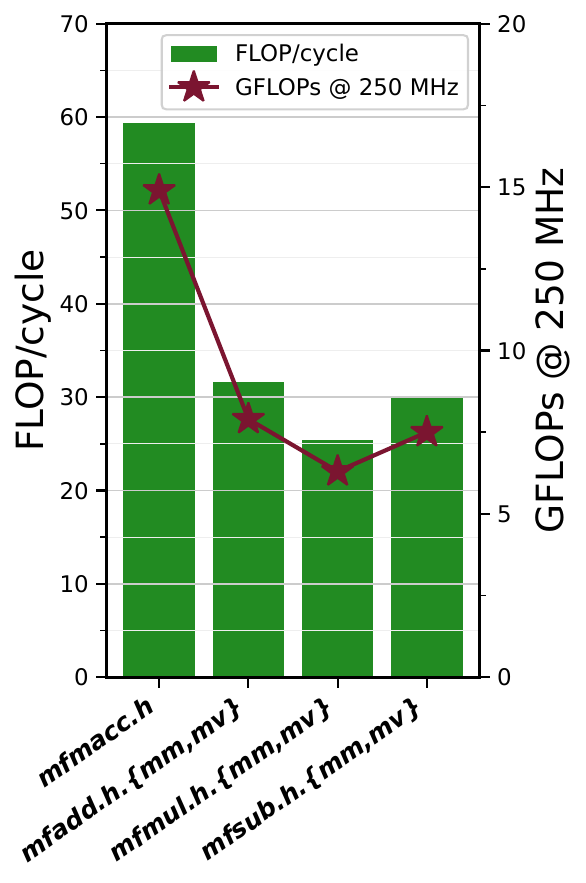}}
    \caption{Relative and absolute performance of AME instructions mapped via PEPs onto HBM-PIM}
    \label{fig:pep_overview}
\end{figure}

\begin{comment}
\begin{table}[h]
\centering
\small
\setlength{\tabcolsep}{10pt}
\renewcommand{\arraystretch}{1.15}
\begin{tabular}{lcc}
\toprule
\textbf{PEP} & \textbf{Setup [cycles]} & \textbf{Exec [cycles]} \\
\midrule
MAC$_{128\times 2048 \cdot 2048\times 1}$   & 90  & 8736 \\
ADD$_{128\times 2048 + 128\times 2048}$    & 82  & 8224 \\
MUL$_{128\times 2048 * 128\times 2048}$    & 84  & 10240 \\
SUB$_{128\times 2048  - 128\times 2048}$    & 82  & 8672 \\
\bottomrule
\end{tabular}
\caption{Cycle counts measured for PIM Execution Primitives. Each primitive is annotated with operand dimensions.}
\label{tab:pim_trigger_results}
\vspace{-4mm}
\end{table}
\end{comment}
%\begin{table*}[ht!]
%\centering
%\small
%\setlength{\tabcolsep}{10pt}
%\renewcommand{\arraystretch}{1.15}
%\begin{tabular}{lcccc}
%\toprule
%\textbf{PEP} 
%& \textbf{MAC$_{128\times 8 \cdot 8\times 1}$} 
%& \textbf{MUL$_{128\times 8 * 128\times 8}$} 
%& \textbf{ADD$_{128\times 8 + 128\times 8}$} 
%& \textbf{SUB$_{128\times 1 - 128\times 1}$} \\
%\midrule
%PIM Setup (cycles) & 82  & 82  & 84  & 82  \\
%PIM Exec (cycles)  & 257 & 320 & 257 & 255 \\
%\bottomrule
%\end{tabular}
%\caption{Cycle counts measured for PIM Execution Primitives. Each primitive is annotated with operand dimensions.}
%\label{tab:pim_trigger_results}
%\end{table*}

%\begin{figure}[h]
    %\centering
    %\includegraphics[width=\linewidth]{plots/pim_cycles_utilization_v2.pdf}
    %\caption{Cycle breakdown per AME instruction (stacked setup/exec) with GFLOP/s on the secondary axis.}
%    \label{fig:pim_cycles_utilization}
%\end{figure}
Overall we can notice that on the maximum tile size the setup cost is lower than the 1\% of the run-time cost which means it can be neglected. 
Applying the previously reported results to AME instructions requires accounting for the number of times each PEP must be invoked to complete a given instruction when operating on maximum-dimension tiles. Specifically, the \emph{mfmacc} instruction requires 256 MAC-PEP invocations, \emph{mfadd}, \emph{mfmul} and \emph{mfsub} require a double invocation. The resulting cycle counts are shown in Figure~\ref{fig:ame-instr-cycles}, along with relative performance in \texttt{FLOP/cycle} and absolute performance in \texttt{GFLOP/s} considering the FPGA frequency of 250MHz (Figure~\ref{fig:ame-instr-gflops}). 
%\vspace{-5mm}
\begin{comment}
\begin{table}[h]
    \centering
    \small
    \setlength{\tabcolsep}{6pt}
    \renewcommand{\arraystretch}{1.05}
    \begin{tabular}{lccc}
        \toprule
        \textbf{AME instr} &
        \textbf{Cycles} &
        \textbf{FLOP/cycle} &
        \textbf{GFLOPs @ 250\,MHz} \\
        \midrule
        \emph{mfmacc.h}              & 2,259,456 & 59.4 & 14.9 \\
        \emph{mfadd.h.\{mm,mv\}}     &    16,616 & 31.6 & 7.9 \\
        \emph{mfmul.h.\{mm,mv\}}     &    20,644 & 25.4 & 6.3 \\
        \emph{mfsub.h.\{mm,mv\}}     &    17,508 & 29.9 & 7.5 \\
        \bottomrule
    \end{tabular}
    \caption{Performance of AME instructions implemented via PEPs on HBM-PIM.}
    \label{tab:ame-perf}
\vspace{-8mm}    
\end{table}
\end{comment}

The maximum theoretical throughput of a pseudo-channel is \texttt{128 FLOP/cycle}. In practice, this performance is not attainable due to the setup phase overhead (for small tile sizes) and the additional operations required to overcome intrinsic limitations of the PIM ISA, such as the lack of cross-lane reduction and native subtraction support.

As reported in Figure~\ref{fig:pep_overview}, the measured \texttt{FLOP/cycle} for the \emph{mfmacc} instruction reflects the ratio between PIM instructions devoted to actual computation within the \texttt{MAC-PEP} and those required for data movement and arrangement. Within the central execution loop, this ratio is approximately 1-to-1, implying that at most half of the theoretical peak performance can be sustained, excluding the initial setup and final data movement overheads. This results in the observed throughput of 59.4~\texttt{FLOP/cycle}. As a comparison, MPC-Wrapper\cite{mpc_wrap} shows a performance of 58.1 \texttt{FLOP/cycle} per single pseudo-channel on a GEMV on a $1024\times128$ matrix. The same principle applies to all other AME instructions implemented. 

To overcome this limitation, we observe for \emph{mfmacc} the need for a single-cycle memory-to-all-bank lane broadcast instruction, which would allow a single operand tile to be distributed across all compute lanes without incurring per-bank load instructions, effectively reducing the data-movement overhead. While such support would require additional interconnect and control complexity, similar broadcast mechanisms are commonly employed in SIMD and GPU architectures, suggesting its feasibility. For element-wise operations, this limitation can be mitigated by fusing multiple instructions (e.g., combining operand loading and arithmetic) or by allowing two bank-resident operands, thereby amortizing memory accesses across multiple computations. These approaches reduce, but do not fully eliminate, the imbalance, as they remain constrained by memory bandwidth and instruction encoding complexity.

\begin{figure}[t]
    \centering
    \includegraphics[width=\linewidth]{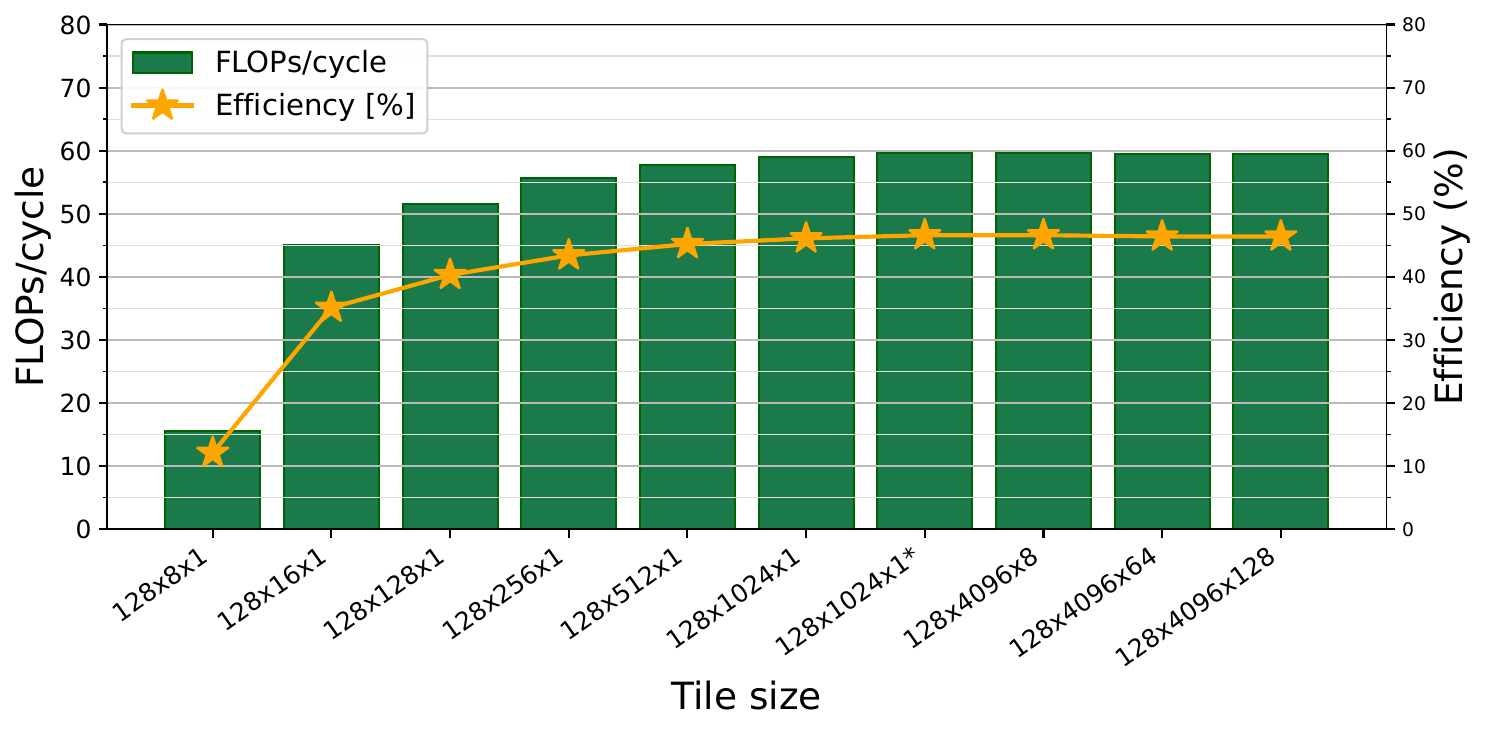}
    \caption{Scaling of FLOPs per cycle for the \emph{mfmacc} instruction with increasing tile size. (*~same performance is obtained by 128$\times$8$\times$256 configuration).}
    \label{fig:efficiency}
\end{figure}

Computational efficiency scales strongly with tile size, as shown in Figure~\ref{fig:efficiency}. 
For smaller tiles, the \texttt{MAC-PEP} requires fewer \texttt{JUMP} iterations than the maximum supported by the PIM ISA, causing setup overheads to dominate. As tile dimensions grow toward $128 \times 2048 \times 1$, efficiency saturates at 59.4 \texttt{FLOP/cycle}, bounded solely by the PIM ISA limitations by the 1-to-1 ratio of compute-to-data-movement instructions within the MAC-PEP loop. The same trend applies to all element-wise PEPs. This scaling behaviour confirms that the proposed HBM-PIM matrix acceleration is best suited to large, tiled dataflows, making it a natural fit for an AME-based implementation. %the execution to be dominated by setup overhead (no jump iterations needed for the tile size $128\times8\times1$). As the number of iterations approaches the architectural limit (e.g., 256 for a matrix--vector operation of $128 \times 2048 \times 1$), the efficiency reaches its maximum and becomes bounded solely by the PIM ISA limitations, such as the absence of cross-lane reduction and the inability to use \texttt{SRF\_Ms} as input operands for the PIM \texttt{MAC} instruction in AAM.
%
%Similar trends are observed for the other element-wise instructions implemented in HBM-PIM, where efficiency decreases with decreasing tile dimensions due to the growing impact of fixed setup costs and underutilized computational resources.
%
%The increasing efficiency of execution as the tile size grows, highlights how the proposed HBM-PIM matrix acceleration is particularly well suited to large, tiled dataflows and therefore to an AME-based implementation.

\section{Related Works}
Prior FPGA-based HBM-PIM accelerators rely on custom designs that retain reduction logic outside memory. The RNN-T accelerator~\cite{rnn_hbm} offloads GEMV to in-memory units but follows a conventional inner-product dataflow, requiring host-side reduction of partial products after write-back to DRAM. MPC-Wrapper~\cite{mpc_wrap} scales performance across 
16 pseudo-channels on Aquabolt-XL but similarly depends on external FPGA logic to aggregate 
intermediate results. Both approaches partially undermine the motivation for PIM by 
introducing significant intermediate data movement. SparsePIM~\cite{sparsepim} and 
IMPRINT~\cite{IMPRINT} extend the HBM-PIM ISA to support irregular workloads, but are 
evaluated through modelling rather than physical implementation, leaving their feasibility 
under memory-compatible logic density constraints unclear. On the software side, the 
programming model shipped with Aquabolt-XL~\cite{hbm_pim_hc} exposes PIM execution as 
raw sequences of DRAM commands and mode transitions, tightly coupling applications to 
device-specific details and limiting portability. Our work differs from all of the above 
by enabling in-memory accumulation for GEMM, GEMV, and element-wise operations on a single 
pseudo-channel through a reduction-free outer-product dataflow, while targeting a portable 
ISA-level abstraction.
\begin{table}[t]
\centering
\caption{Comparison of AME-PIM with related works.}
\label{tab:comparison}
\resizebox{\columnwidth}{!}{%
\begin{tabular}{l c c c c c c}
\toprule
\rotatebox{60}{\textbf{Work}} &
\rotatebox{60}{\shortstack{Pseudo-channels\\(pCH)}} &
\rotatebox{60}{In-memory acc.} &
\rotatebox{60}{GEMM} &
\rotatebox{60}{GEMV} &
\rotatebox{60}{Eltwise Ops} &
\rotatebox{60}{\shortstack{GEMM/GEMV\\FLOP/cycle per pCH}}\\
\midrule
RNN-T [9]         & 1             & \xmark & \xmark & \cmark & \xmark & n.a.\\
MPC-Wrapper [4]   & 16            & \xmark & \cmark & \cmark & \xmark & 58.1\\
\midrule
\textbf{This work} & \textbf{1} & \cmark & \cmark & \cmark & \cmark & 59.4\\
\bottomrule
\end{tabular}%
}
\vspace{-2mm}
\end{table}
\section{Conclusions and Future Work}
With the objective of addressing the \emph{von Neumann bottleneck}, this work analysed the capabilities of HBM-PIM with the objective of employing it as a hardware substrate for the implementation of an ISA-level matrix extension. Using the RISC-V AME T-HEAD proposal as a target, we investigated whether the majority of matrix computations could be executed directly within HBM-PIM, without relying on external computational engines and dedicated software stacks. For most operations, this approach proved feasible, ranging from element-wise arithmetic to matrix tile multiplication. We proposed mappings that rely on specific dataflows and architectural workarounds to compensate for missing features in the PIM execution model, such as native reduction and subtraction support. However, certain operations, including element-wise maximum and minimum, could not be implemented under these constraints.

We conducted a preliminary performance evaluation, indicating promising performance and scaling for matrix tile multiplication, while also highlighting the significant computational overhead required to overcome PIM ISA restrictions. These results emphasize both the potential and the current limitations of using HBM-PIM as a backend for an ISA-level matrix extension.

Future work will focus on a more comprehensive performance evaluation of the proposed programming model, including the cost of data movement instructions and the coordinated use of all available HBM-PIM pseudo-channels. The ultimate goal is to achieve a fully functional AME implementation that fully exploits the computational capabilities of the HBM-PIM platform.

\begin{acks} 
This activity has been supported by the DARE (g.a. 101143421) project and the ICSC Spoke future HPC. The authors gratefully acknowledge Samsung for the long-term loan of modified Xilinx Alveo U280 platforms with HBM-PIM support used in this work.
\end{acks}
\bibliographystyle{ACM-Reference-Format}
\bibliography{sample-base}

@String{Computing = "Computing" }

@String{Computer = "{IEEE} Computer" }

@INPROCEEDINGS{PIM-power,
  author={Kwon, Young-Cheon and Lee, Suk Han and Lee, Jaehoon and Kwon, Sang-Hyuk and Ryu, Je Min and Son, Jong-Pil and Seongil, O and Yu, Hak-Soo and Lee, Haesuk and Kim, Soo Young and Cho, Youngmin and Kim, Jin Guk and Choi, Jongyoon and Shin, Hyun-Sung and Kim, Jin and Phuah, BengSeng and Kim, HyoungMin and Song, Myeong Jun and Choi, Ahn and Kim, Daeho and Kim, SooYoung and Kim, Eun-Bong and Wang, David and Kang, Shinhaeng and Ro, Yuhwan and Seo, Seungwoo and Song, JoonHo and Youn, Jaeyoun and Sohn, Kyomin and Kim, Nam Sung},
  booktitle={2021 IEEE International Solid-State Circuits Conference (ISSCC)}, 
  title={25.4 A 20nm 6GB Function-In-Memory DRAM, Based on HBM2 with a 1.2TFLOPS Programmable Computing Unit Using Bank-Level Parallelism, for Machine Learning Applications}, 
  year={2021},
  volume={64},
  number={},
  pages={350-352},
  doi={10.1109/ISSCC42613.2021.9365862}}

@INPROCEEDINGS{aquabolt,
  author={Lee, Sukhan and Kang, Shin-haeng and Lee, Jaehoon and Kim, Hyeonsu and Lee, Eojin and Seo, Seungwoo and Yoon, Hosang and Lee, Seungwon and Lim, Kyounghwan and Shin, Hyunsung and Kim, Jinhyun and Seongil, O and Iyer, Anand and Wang, David and Sohn, Kyomin and Kim, Nam Sung},
  booktitle={2021 ACM/IEEE 48th Annual International Symposium on Computer Architecture (ISCA)}, 
  title={Hardware Architecture and Software Stack for PIM Based on Commercial DRAM Technology : Industrial Product}, 
  year={2021},
  volume={},
  number={},
  pages={43-56},
  doi={10.1109/ISCA52012.2021.00013}}

@misc{JEDEC:2013:HBM,
  author = {{JEDEC Solid State Technology Association}},
  title = {{High Bandwidth Memory (HBM) DRAM}},
  howpublished = {JESD235},
  year = {2013},
  note = {Standard document, available from \url{www.jedec.org}},
  url = {www.jedec.org}
}

@article{jia2018dissecting,
  title={Dissecting the NVIDIA volta GPU architecture via microbenchmarking},
  author={Jia, Zhe and Maggioni, Marco and Staiger, Benjamin and Scarpazza, Daniele P},
  journal={arXiv preprint arXiv:1804.06826},
  year={2018}
}

@INPROCEEDINGS{hbm_si,
  author={Cho, Kyungjun and Lee, Hyunsuk and Kim, Joungho},
  booktitle={2016 Pan Pacific Microelectronics Symposium (Pan Pacific)}, 
  title={Signal and power integrity design of 2.5D HBM (High bandwidth memory module) on SI interposer}, 
  year={2016},
  volume={},
  number={},
  pages={1-5},
  doi={10.1109/PanPacific.2016.7428425}}

@INPROCEEDINGS{HBM_HC,
  author={Joonyoung Kim and Younsu Kim},
  booktitle={2014 IEEE Hot Chips 26 Symposium (HCS)}, 
  title={HBM: Memory solution for bandwidth-hungry processors}, 
  year={2014},
  volume={},
  number={},
  pages={1-24},
  doi={10.1109/HOTCHIPS.2014.7478812}}

@ARTICLE{HBM_DRAM,
  author={Sohn, Kyomin and Yun, Won-Joo and Oh, Reum and Oh, Chi-Sung and Seo, Seong-Young and Park, Min-Sang and Shin, Dong-Hak and Jung, Won-Chang and Shin, Sang-Hoon and Ryu, Je-Min and Yu, Hye-Seung and Jung, Jae-Hun and Lee, Hyunui and Kang, Seok-Yong and Sohn, Young-Soo and Choi, Jung-Hwan and Bae, Yong-Cheol and Jang, Seong-Jin and Jin, Gyoyoung},
  journal={IEEE Journal of Solid-State Circuits}, 
  title={A 1.2 V 20 nm 307 GB/s HBM DRAM With At-Speed Wafer-Level IO Test Scheme and Adaptive Refresh Considering Temperature Distribution}, 
  year={2017},
  volume={52},
  number={1},
  pages={250-260},
  doi={10.1109/JSSC.2016.2602221}}

@INPROCEEDINGS{hbm_pim_hc,
  author={Kim, Jin Hyun and Kang, Shin-haeng and Lee, Sukhan and Kim, Hyeonsu and Song, Woongjae and Ro, Yuhwan and Lee, Seungwon and Wang, David and Shin, Hyunsung and Phuah, Bengseng and Choi, Jihyun and So, Jinin and Cho, YeonGon and Song, JoonHo and Choi, Jangseok and Cho, Jeonghyeon and Sohn, Kyomin and Sohn, Youngsoo and Park, Kwangil and Kim, Nam Sung},
  booktitle={2021 IEEE Hot Chips 33 Symposium (HCS)}, 
  title={Aquabolt-XL: Samsung HBM2-PIM with in-memory processing for ML accelerators and beyond}, 
  year={2021},
  volume={},
  number={},
  pages={1-26},
  doi={10.1109/HCS52781.2021.9567191}}

@INPROCEEDINGS{mpc_wrap,
  author={Choi, Jinwoo and Ha, Yeonan and Cha, Hanna and Lee, Seil and Lee, Sungchul and Lee, Jounghoo and Kang, Shin-haeng and Kim, Bongjun and Jung, Hanwoong and Kim, Hanjun and Kim, Youngsok},
  booktitle={2024 IEEE 32nd Annual International Symposium on Field-Programmable Custom Computing Machines (FCCM)}, 
  title={MPC-Wrapper: Fully Harnessing the Potential of Samsung Aquabolt-XL HBM2-PIM on FPGAs}, 
  year={2024},
  volume={},
  number={},
  pages={162-172},
  doi={10.1109/FCCM60383.2024.00027}}

@inproceedings{quadrilatero,
author = {Cammarata, Danilo and Perotti, Matteo and Bertuletti, Marco and Garofalo, Angelo and Schiavone, Pasquale Davide and Atienza, David and Benini, Luca},
title = {Quadrilatero: A RISC-V programmable matrix coprocessor for low-power edge applications},
year = {2025},
isbn = {9798400713934},
publisher = {Association for Computing Machinery},
address = {New York, NY, USA},
url = {https://doi.org/10.1145/3706594.3726978},
doi = {10.1145/3706594.3726978},
booktitle = {Proceedings of the 22nd ACM International Conference on Computing Frontiers: Workshops and Special Sessions},
pages = {66–69},
numpages = {4},
location = {
},
series = {CF '25 Companion}
}

@inproceedings{rnn_hbm,
author = {Kang, Shinhaeng and Lee, Sukhan and Kim, Byeongho and Kim, Hweesoo and Sohn, Kyomin and Kim, Nam Sung and Lee, Eojin},
title = {An FPGA-based RNN-T Inference Accelerator with PIM-HBM},
year = {2022},
isbn = {9781450391498},
publisher = {Association for Computing Machinery},
address = {New York, NY, USA},
url = {https://doi.org/10.1145/3490422.3502355},
doi = {10.1145/3490422.3502355},
booktitle = {Proceedings of the 2022 ACM/SIGDA International Symposium on Field-Programmable Gate Arrays},
pages = {146–152},
numpages = {7},
location = {Virtual Event, USA},
series = {FPGA '22}
}

@misc{t_head_ame_v0_6_0,
  title        = {{RISC-V Matrix Specification Proposal}},
  author       = {{XuanTie RISC-V Team}},
  howpublished = {\url{https://github.com/XUANTIE-RV/riscv-matrix-extension-spec}},
  note         = {Version v0.6.0, 2024-12-11: Draft},
  year         = {2024}
}

@misc{xuantie_c907_ame,
  title        = {{Enhancing the Future of AI/ML with Attached Matrix Extension}},
  author       = {{RISC-V International}},
  howpublished = {\url{https://riscv.org/blog/enhancing-the-future-of-ai-ml-with-attached-matrix-extension/}},
  year         = {2024},
  note         = {Accessed: 2026-01-22}
}

@misc{intel_ise_prm_amx,
  title        = {{Intel\textsuperscript{\textregistered} Architecture Instruction Set Extensions and Future Features Programming Reference}},
  author       = {{Intel Corporation}},
  howpublished = {\url{https://cdrdv2-public.intel.com/836496/architecture-instruction-set-extensions-programming-reference.pdf}},
  year         = {2023},
  note         = {Includes Intel AMX architectural overview and instructions. Accessed: 2026-01-26}
}

@misc{arm_sme_intro,
  title        = {{Arm Scalable Matrix Extension (SME) Introduction}},
  author       = {{Arm Ltd.}},
  howpublished = {\url{https://developer.arm.com/community/arm-community-blogs/b/architectures-and-processors-blog/posts/arm-scalable-matrix-extension-introduction}},
  year         = {2024},
  note         = {Accessed: 2026-01-26}
}

@inproceedings{IMPRINT,
author = {Cukurtas, Ersin and Ranawella, Kavish and Skadron, Kevin and Stan, Mircea},
title = {IMPRINT: In-Memory Processing with Indirect Addressing Techniques for GPU-hosted HBM-PIM},
year = {2026},
isbn = {9798400720024},
publisher = {Association for Computing Machinery},
address = {New York, NY, USA},
url = {https://doi.org/10.1145/3767110.3767121},
doi = {10.1145/3767110.3767121},
booktitle = {Proceedings of the International Symposium on Memory Systems},
pages = {165–176},
numpages = {12},
location = {
},
series = {MemSys '25}
}

@inproceedings{sparsepim,
author = {Kang, Taewoon and Choi, Geonwoo and Suh, Taeweon and Koo, Gunjae},
title = {SparsePIM: An Efficient HBM-Based PIM Architecture for Sparse Matrix-Vector Multiplications},
year = {2025},
isbn = {9798400715372},
publisher = {Association for Computing Machinery},
address = {New York, NY, USA},
url = {https://doi.org/10.1145/3721145.3735111},
doi = {10.1145/3721145.3735111},
booktitle = {Proceedings of the 39th ACM International Conference on Supercomputing},
pages = {495–512},
numpages = {18},
location = {
},
series = {ICS '25}
}

%\appendix

\end{document}